\documentclass[nofootinbib,amsmath,amssymb,twocolumn,superscriptaddress]{revtex4-2}
\usepackage[english]{babel}
\usepackage{lmodern}
\usepackage{microtype}
\usepackage{booktabs}
\usepackage{bm}
\usepackage{graphicx}
\usepackage{listings}
\usepackage{physics}
\usepackage{subfigure}
\usepackage{fancyhdr}
\usepackage{xcolor}
\usepackage[hidelinks]{hyperref}
\usepackage{natbib}
\usepackage{soul}

\newcommand{\orcidicon}[1]{\href{https://orcid.org/#1}{\includegraphics[width=8pt]{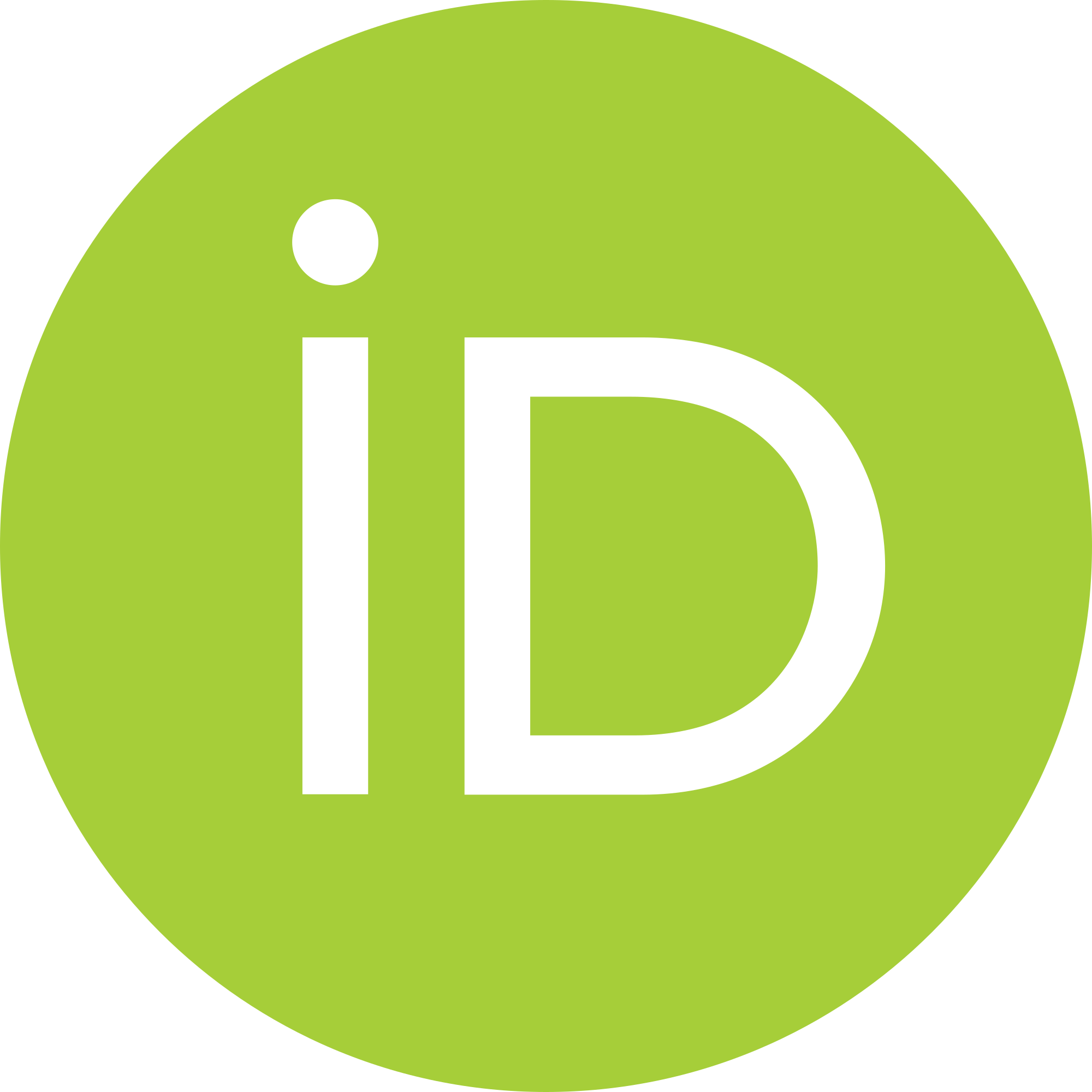}}}
\newcommand{\orcid}[1]{\,\href{https://orcid.org/#1}{\protect\orcidicon{#1}}}

\newcommand{\mnras}{Mon. Notices Royal Astron. Soc.}
\newcommand{\apjl}{Astrophys. J. Lett.}
\newcommand{\aap}{Astron. Astrophys.}
\newcommand{\prx}{Phys. Rev. X}
\newcommand{\cqg}{Class. Quant. Grav.}

\newcommand{\msun}{\mathrm{M}_\odot}

\begin{document}
\title{GW190521: a binary black hole merger inside an active galactic nucleus?} 
\author{Sophia~L.~Morton\orcid{0000-0002-2658-3430}}
\email{mortonso@oregonstate.edu}
\affiliation{Department of Physics, Oregon State University, Corvallis, OR 97331, USA}
\author{Stefano Rinaldi\orcid{0000-0001-5799-4155}}
\email{stefano.rinaldi@uni-heidelberg.de}
\affiliation{Dipartimento di Fisica ``E. Fermi'', Università di Pisa, I-56127 Pisa, Italy}
\affiliation{INFN, Sezione di Pisa, I-56127 Pisa, Italy}
\author{Alejandro Torres-Orjuela\orcid{0000-0002-5467-3505}}
\email{atorreso@hku.hk}
\affiliation{Department of Physics, The University of Hong Kong, Pokfulam Road, Hong Kong}
\author{Andrea Derdzinski\orcid{0000-0001-9880-8929}}
\affiliation{Department of Life and Physical Sciences, Fisk University, 1000 17th Avenue N., Nashville, TN 37208, USA}
\affiliation{Department of Physics \& Astronomy, Vanderbilt University,
2301 Vanderbilt Place, Nashville, TN 37235, USA}
\author{M. Paola Vaccaro\orcid{0000-0003-3776-9246}}
\affiliation{Institut f{\"u}r Theoretische Astrophysik, ZAH, Universit{\"a}t Heidelberg, Albert-Ueberle-Stra{\ss}e 2, D-69120 Heidelberg, Germany} 
\author{Walter Del Pozzo\orcid{0000-0003-3978-2030}}
\affiliation{Dipartimento di Fisica ``E. Fermi'', Università di Pisa, I-56127 Pisa, Italy}
\affiliation{INFN, Sezione di Pisa, I-56127 Pisa, Italy}
\date{\today}

\begin{abstract}
GW190521, the most massive binary black hole merger confidently detected by the LIGO-Virgo-KAGRA collaboration, is the first gravitational-wave observation of an intermediate-mass black hole. The signal was followed approximately 34 days later by flare ZTF19abanrhr, detected in AGN J124942.3+344929 by the Zwicky Transient Facility at the 78\% spatial contour for GW190521’s sky localization. Using the GWTC-2.1 data release, we find that the association between GW190521 and flare ZTF19abanrhr as its electromagnetic counterpart is preferred over a random coincidence of the two transients with a log Bayes’ factor of 8.6, corresponding to an odds ratio of $\sim 5400:1$ for equal prior odds and $\sim 400:1$ assuming an astrophysical prior odds of 1/13. Given the association, the multi-messenger signal allows for an estimation of the Hubble constant, finding $H_0 = 102^{+27}_{-25}\mathrm{\ km \ s^{-1} \ Mpc^{-1}}$ when solely analyzing GW190521 and $79.2^{+17.6}_{-9.6}\mathrm{\ km \ s^{-1} \ Mpc^{-1}}$ assuming prior information from the binary neutron star merger GW170817, both consistent with the existing literature.
\end{abstract}

\maketitle

\section{Introduction}\label{introduction}
On May 21, 2019, the two LIGO interferometers \citep{ligodetector:2015} and the Virgo interferometer \citep{virgodetector:2015} detected GW190521, a reported binary black hole (BBH) merger between a $85\ \msun$ and a $66 \ \msun$ black hole (BH) that produced a $142\ \msun$ BH \cite[hereafter also \emph{discovery paper}] {GW190521:discovery}, potentially the first intermediate-mass BH confidently observed. GW190521's high primary mass $M_1$ was found in tension with the earlier population inferred with the observations from the first and second LIGO-Virgo-KAGRA (LVK) Observing Runs (O1 and O2) \citep{astrodistGWTC1:2019}. Subsequent population studies including the third Observing Run (O3) observations, however, reconciled this event with the rest of the BH population for some models \citep{astrodistGWTC2:2021,astrodist:gwtc3}.

The main model used in \citet{astrodist:gwtc3}, the \textsc{PowerLaw+Peak} model, is inspired by the initial stellar mass function \citep{salpeter:1955,kroupa:2001} and revolves around a power-law model with the addition of a Gaussian peak and a high mass cutoff. The latter feature is included to account for the presence of a mass gap at large masses due to the inability of astrophysical processes to form black holes in that range \citep{GW190521:implications,belczynski:2016,fowler:1964,marchant:2019,spera:2017,stevenson:2019,woosley:2017,woosley:2021}.
\citet{astrodist:gwtc3} finds that the low-mass end of the mass gap is above 75 $\msun$ at 96.9\% credibility, while inferring the population properties of the BHs observed up to O3.
Since the primary mass of GW190521 is on the edge of the mass gap, several other possible explanations have been proposed to interpret the existence of this massive system, such as the primary mass coming from the far side of the mass gap \citep{fishbach:2020:gw190521}, introducing a large orbital eccentricity \citep{gayathri:2020,romeroshaw:2020}, formation through dynamical capture \citep{gamba:2023}, or new physics \citep{sakstein:2020,wang:2022}.

Approximately 34 days after GW190521, the Zwicky Transient Facility (ZTF) observed a flare, ZTF19abanrhr, in an active galactic nucleus (AGN) located within the spatial footprint of the detected GW event \citep{graham:2020}. The AGN in question, AGN J124942.3+344929, is reported by \citet{graham:2020} to have a very consistent luminosity for the year surrounding the event, making it unlikely that flare ZTF19abanrhr is due to the intrinsic variability of the source. Moreover, the observation of the flare showed little color variation, suggesting its cause to be a constant temperature shock to the AGN gas.

BBHs in AGNs are expected to produce electromagnetic (EM) counterparts   as a result of accretion onto the remnant BH that produces jet launching{, outflow bubbles, breakout emission, and shock cooling \citep{Kimura:2021,Rodriguez-Ramirez:2023,TagawaHiromichi:2023,tagawa:2023}.} 
Additionally, the collisional interaction between the merger-remnant BH and the surrounding gas can produce shocks \citep{mckernan:2019,graham:2023},
as a result of the kick velocity induced by the significant loss of linear momentum from GWs \citep{GW190521:implications}.
{Each of these mechanisms produces delayed X-ray, optical, UV, or infrared flares} on top of the AGN emission.
For the GW190521 detection, the ZTF identified a possible counterpart (ZTF19abanrhr) applying this model \citep{graham:2020} 
and the observations are consistent with these predictions. 

In the accretion disks of AGNs the BH mass gap is also not expected to be present due to the increased probability for hierarchical mergers and accretion \citep{vaccaro:2023}.
In this formation model, BHs can either form within the disk in gravitationally unstable regions \citep{stone:2017,derdzinski:2023} or be captured into the disk plane via drag forces \citep{syer:1991,tagawa:2020}. Subsequent migration through the disk prompts mass accretion \citep{mckernan:2012,yang:2019}, and can lead to BH accumulation in the migration trap \citep{bellovary:2016}.
The deep potential well of the central supermassive black hole (SMBH) not only facilitates high-mass BH mergers but also influences the evolution of the binary and its associated GW signal, as we discuss in this work.
Conversely, the presence of matter in the surroundings of the binary system seems to have a negligible effect on the GW itself \citep{canevasantoro:2023,leong:2023}.

The association between the AGN flare and the GW event has been originally discussed in \citet{ashton:2021}, where the authors find marginal evidence in favor of the BBH and the EM counterpart having the same origin, mainly due to the poor overlap between the discovery paper luminosity distance posterior distribution and the AGN's redshift. The LVK collaboration released an updated set of posterior samples under the name of GWTC-2.1 \citep{gwtc2.1}, where the events detected up to the first half of O3 (O3a) were re-analyzed with improved calibration and noise subtraction and updated waveforms with respect to the initial catalog GWTC-2 \citep{gwtc2:2021}.
The most recent data set for GW190521 includes a substantially different posterior distribution for the luminosity distance $D_L$, placing the BBH significantly closer to us than previously thought. This opened up the possibility for a different answer to the question \emph{is there an association between GW190521 and the EM counterpart candidate?}

The rest of this paper is outlined as follows: we first discuss the motivation for re-investigating the association between GW190521 and EM flare ZTF19abanrh in Sec. \ref{motivation}. Then we present a theoretical background for our model in which the BBH resides in an AGN and additional redshift is present in Sec. \ref{Framework}. The results in Sec. \ref{Results: association model} compare our association model to the coincidence model in which the GW and EM events are unrelated. Additionally, in Sec.~\ref{Hubble Constant} the Hubble constant is estimated within an association model both with and without prior information from the multi-messenger event GW170817.

\section{Motivation}\label{motivation}
\begin{figure} 
  \begin{center}
    \includegraphics[width=\columnwidth]{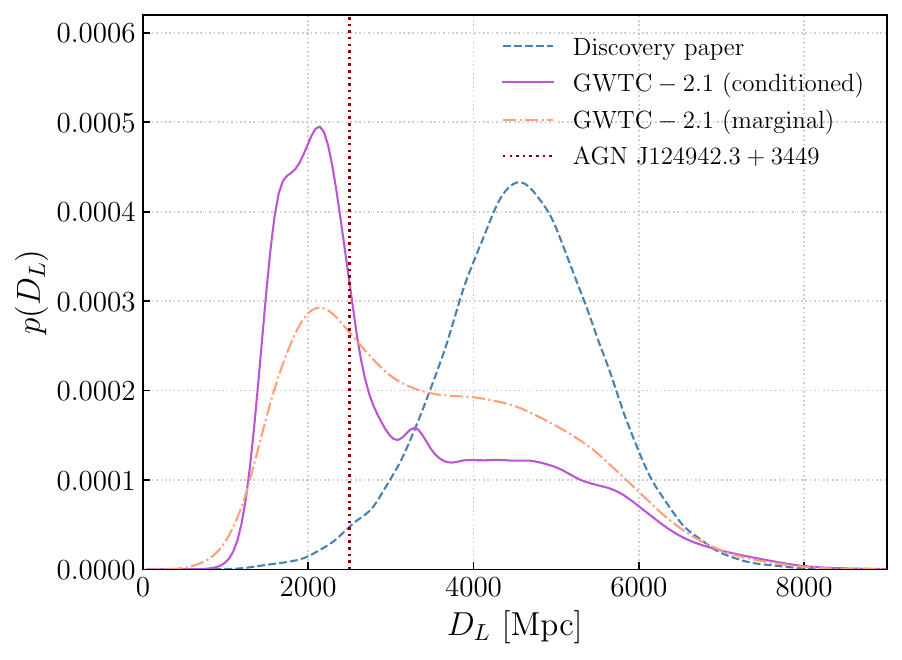}
  \end{center}
  \caption{Posteriors on luminosity distance using the IMRPhenomPv3HM waveform from \citet{GW190521:discovery} (dashed blue), GWTC-2.1 posterior marginalized over sky position (dot-dashed orange), and conditioned on the sky position of AGN J124942.3+3449 (solid purple). The AGN distance is shown as a dashed line in red for comparison. The older data release prefers further distances than GWTC-2.1 data.}
  \label{fig:D_L}
\end{figure}

When analyzing the odds of association between two astrophysical events, a significant contribution comes from the three dimensional spatial overlap. In this case, the GW luminosity distance posteriors can be compared to the luminosity distance of the EM flare, obtained from the AGN's redshift.
Prior works by \citet{ashton:2021} {and \citet{palmese:2021}} investigate the possible association between GW190521 and ZTF19abanrhr to have a common origin using the initial LVK data release associated with the discovery paper \citep{GW190521:discovery} and find only marginal evidence in favor of the association. The original data indicate a poor luminosity distance overlap between the GW source and the EM counterpart, with the AGN lying at $1.6\pm0.7\ \sigma$ \citep{ashton:2021}from the GW peak marginal luminosity distance of $5.3^{+0.30}_{-0.11}\ \mathrm{Gpc}$ \citep{graham:2020} (see Fig.~\ref{fig:D_L}).

The luminosity distance discrepancy between the GW and EM events with the discovery paper posterior samples \citep{GW190521:discovery} led to the inability to confidently associate GW190521 with the EM candidate. In \citet{ashton:2021} with the discovery paper data, odds ratios were only 1 in 12 or less -- depending on the waveform model used -- in favor of the association. The Bayes' factors for the common origin hypothesis versus coincidence are mainly driven by the spatial posterior overlap integrals for each waveform presented in \citet{GW190521:discovery}. In their work, due to the better precision of EM measurements with respect to GW posterior distributions, the sky position and luminosity distance for the source are taken to be that of the AGN, neglecting uncertainties: here we make use of the same approximation.
For the IMRPhenomPv3HM waveform, as compared throughout this work, the odds in favor of the common source hypothesis is 2 when assuming that the luminosity distance and sky position are independent.

In a more recent data release, GWTC-2.1, more robust waveforms provide significantly different mass and luminosity distance posteriors for GW190521. The GWTC-2.1 posterior samples \citep{gwtc2.1} favor closer luminosity distances of $3.3^{+2.8}_{-1.8} \ \rm Gpc$.
This is significantly closer to the AGN distance of $\sim$ 2.5 Gpc compared to \citet{GW190521:discovery}, as shown in Fig. \ref{fig:D_L}: the EM counterpart distance lies at the 31$^{\rm st}$ percentile for the distribution marginalized over the entire sky position of the GW source, as opposed to the discovery paper posterior distribution (2$^{\rm nd}$ percentile). When the GW luminosity distance posterior distribution is conditioned on the sky location of AGN J124942.3+3449\footnote{The conditioned posterior distribution is obtained making use of the mathematical properties of the Dirichlet process Gaussian mixture model, as discussed in Section~\ref{sec:codes}.}, the AGN lies at the 49$^{\rm th}$ percentile of the GW luminosity distance posterior. This significant change in the spatial overlap motivates the need for further analysis of the association given that a dominant factor in determining the odds of association is based on the luminosity distance.

\section{Methods}
\label{Framework}
In this section, we will discuss the effects of a SMBH's gravitational well on the GW signal emitted by a nearby merging BBH and present the Bayesian framework used in this analysis.

\subsection{Environmental effects}
\label{Environmental effects}
A BBH residing in an AGN disk is influenced by different environmental effects that impact the properties of the source and the parameter estimation of the binary. Due to the proximity of the BBH to the SMBH in the center of the AGN disk, the observed mass and distance of the source are altered. The motion of the BBH when orbiting the SMBH induces a relativistic redshift
\begin{equation}\label{eq:relrs}
    z_{\rm rel} = \gamma(1+v\cos(\theta)) - 1\,,
\end{equation}
where $\gamma := (1-v^2)^{-1/2}$ is the Lorentz factor, $v$ is the magnitude of the velocity, and $\theta$ is the viewing angle between the velocity and the line of sight in the observer frame. Assuming that the BBH is on a circular orbit around a non-spinning SMBH~\citep{fabj_nasim_2020}, its velocity can be calculated as
\begin{equation}
    v = \frac{1}{\sqrt{2\qty(r/R_s -1)}}\,,
\end{equation}
where $R_s$ is the Schwarzschild radius of the SMBH and $r$ is the distance between the BBH and the SMBH. Furthermore, the gravitational potential of the non-spinning SMBH induces a gravitational redshift
\begin{equation}\label{eq:grars}
    z_{\rm grav} = \sqrt{1-\frac{R_s}{r}} - 1\,.
\end{equation}
Note that SMBHs in AGN disks are expected to be spinning~\citep{king_pringle_2008}. However, we do not include the effect of the spin because, for the distances we expect the BBH to be with respect to the SMBH, it becomes negligible \citep{liu_lai_2019,fang_huang_2019,chen_zhang_2022}.

For a redshifted source, the detected mass differs from the source frame intrinsic mass by a factor equal to the product of all redshifts~\citep{chen_li_2019}
\begin{equation}\label{eq:rsmass}
    M_1^{z,\mathrm{eff}} = (1+z_{\rm c})(1+z_{\rm rel})(1+z_{\rm grav})M_1\,,
\end{equation}
where $z_{\rm c}$ is the cosmological redshift, and $M_1^{z,\mathrm{eff}}$ and $M_1$ are the observed (detector-frame) and intrinsic (source-frame) mass of the source, respectively. The effective distance of the source is also affected by the redshifts but with a square factor for the relativistic redshift term induced by aberration of GWs~\citep{torresorjuela:2023},
\begin{equation}\label{eq:rsdist}
    D_L^{\rm eff} = (1+z_{\rm rel})^2(1+z_{\rm grav})D_L\,,
\end{equation}
where $D_L=(1+z_{\rm c})D_{\rm com}$ is the luminosity distance of the source and $D_{\rm com}$ is the comoving distance between source and observer. In summary, the environmental effects of the SMBH can be accounted for by simply replacing $M_1^z$ with $M_1^{z,\mathrm{eff}}$ and $D_L$ with $D_L^\mathrm{eff}$.

We point out that, when detecting higher spherical modes, the velocity of the source can in principle be detected removing the ambiguity between the relativistic redshift and the other redshifts \citep{gualtieri_berti_2008,boyle:2016,torresorjuela:2021}. However, when detection is mainly done with the dominant quadrupolar mode, as it is the case here, the degeneracy between mass and redshift remains unresolved \citep{yan_chen_2023}.

\subsection{Bayesian statistical framework}\label{Bayesian statistical framework}

In order to assess whether it is more likely that the GW event and the EM transient share a common origin or they are associated by chance, we work within a Bayesian statistical framework: thus, our figure of merit will be the odds ratio $\mathcal{O}^A_C$ between the two models,
\begin{equation}
    \mathcal{O}^A_C= \frac{p(\mathcal{H}_A | d)}{p(\mathcal{H}_C | d)}=\frac{p(d|\mathcal{H}_A)}{p(d|\mathcal{H}_C)}\frac{p(\mathcal{H}_A)}{p( \mathcal{H}_C)} = \mathcal{B}^A_C\mathcal{P}^A_C\,.
\end{equation}
Here $d$ is the data and $\mathcal{H}_A$ and $\mathcal{H}_C$ are the competing hypothesis A and C (association model and coincidence model respectively). 
Following the Bayesian principle of conditional probability, the odds ratio $\mathcal{O}^A_C$ combines our previous knowledge or belief on the analyzed phenomenon, conveyed via the so-called \emph{prior odds} $\mathcal{P}^A_C$, with the information carried by the available data, included in the Bayes' factor,
\begin{equation}
    \mathcal{B}^A_C =  \frac{p(d | \mathcal{H}_A)}{p(d | \mathcal{H}_C)}\,. 
\end{equation}
In most cases, there is no reason to believe that one model is a priori more likely than the other, thus it is often assumed that $\mathcal{P}^A_C = 1$ and the model selection is performed on the Bayes' factor only.
The problem that we are addressing here, however, is one of the few cases in which it is possible to compute a prior probability for the two competing models: in fact, \citet{ashton:2021} uses an astrophysical prior odds of 1/13.

The Bayes' factor $\mathcal{B}^A_C$ is the ratio of the evidences for the two competing models. For a single model, this quantity can be computed via the marginalization over all parameters $\lambda$ required by our model $\mathcal{H}_i$:
\begin{equation}\label{eq:evidence_definition}
p(d|\mathcal{H}_i) = \int p(d |\lambda,\mathcal{H}_i)p(\lambda|\mathcal{H}_i)\dd \lambda \,.
 \end{equation}
The parameters of interest for this work are the source luminosity distance $D_L$, the detector-frame primary mass $M_1^z$, and the sky position of the host galaxy in right ascension $\alpha$ and declination $\delta$: 

\begin{multline}
p(d|\mathcal{H}_i) = \int p(d |M_1^z,D_L,\alpha,\delta,\mathcal{H}_i)\times \\ \times p(M_1^z,D_L,\alpha,\delta|\mathcal{H}_i)\dd M_1^z \dd D_L \dd \alpha \dd \delta \,.
 \end{multline}
Making use of Bayes' theorem we can write 
\begin{multline}\label{eq:likelihood_generic}
p(d|\mathcal{H}_i) = \int \frac{p(M_1^z,D_L,\alpha,\delta|d)}{p(M_1^z,D_L,\alpha,\delta)} \times \\ \times p(M_1^z,D_L,\alpha,\delta|\mathcal{H}_i)\dd M_1^z \dd D_L \dd \alpha \dd \delta\,.
 \end{multline}
In particular, $p(M_1^z,D_L,\alpha,\delta|d)$ is the posterior probability density for the parameters of the GW and $p(M_1^z,D_L,\alpha,\delta)$ is the prior probability on the same quantities used during the parameter estimation (PE) run by the LVK collaboration. In particular, the PE prior is uniform in primary mass, right ascension and declination, and follows \citet{farr:prior} for the luminosity distance:
\begin{equation}
\frac{p(M_1^z,D_L,\alpha,\delta|d)}{p(M_1^z,D_L,\alpha,\delta)} \propto \frac{p(M_1^z,D_L,\alpha,\delta|d)}{p(D_L)}\,.
\end{equation}

The second term of Eq.~\eqref{eq:likelihood_generic}, $p(M_1^z,D_L,\alpha,\delta|\mathcal{H}_i)$, is the prior conditioned on the hypothesis we are considering. Under the assumption that these quantities are a priori independent, we can factorize it as:
\begin{equation}\label{eq:factorised_prior}
    p(M_1^z,D_L|\mathcal{H}_i)p(\alpha,\delta|\mathcal{H}_i)\,.
\end{equation}
The following sections specify the prescriptions for these distributions under the two competing models (or hypotheses) considered in this work.

\subsubsection{Association Model}
In the association model $\mathcal{H}_A$, flare ZTF19abanrh and GW190521 share a common origin and location. The sky location is thus fixed to the AGN location $(\alpha_\mathrm{AGN}$, $\delta_\mathrm{AGN})$, and the prior for the sky position reads:
\begin{equation}
    p(\alpha,\delta|\mathcal{H}_A) = \delta\qty(\alpha-\alpha_\mathrm{AGN})\delta\qty(\delta-\delta_\mathrm{AGN})\,.
\end{equation}
As described in Sec.~\ref{Environmental effects}, the parameters that enter in the GW posterior probability density under the common origin hypothesis used in this work are the detector-frame effective primary mass $M^{z,\mathrm{eff}}_1$ and effective luminosity distance $D_L^\mathrm{eff}$. 
These quantities depend on the additional redshift introduced by environmental effects dependent on the distance from the SMBH $r$, the orbital position $\theta$, and the cosmological redshift $z_\mathrm{c}$, which is obtained from the EM source\footnote{\url{https://skyserver.sdss.org/dr12/en/tools/explore/Summary.aspx?id=1237665128546631763}}: $z_\mathrm{AGN} = 0.438$. The prior then reads:
\begin{multline}
    \label{marginalization} p(M_1^{z,\mathrm{eff}},D_L^\mathrm{eff}|\mathcal{H}_A) = \int p(M_1^{z,\mathrm{eff}},D_L^\mathrm{eff}|z_\mathrm{c}, r, \theta, M_1, \mathcal{H}_A)\times\\ \times p(z_\mathrm{c}|\mathcal{H}_A)p(r|\mathcal{H}_A)p(\theta|\mathcal{H}_A)p(M_1|\mathcal{H}_A)\dd z_c \dd r \dd \theta \dd M_1\,,
\end{multline}
where we assumed the independence of the prior distributions.
The first term of the integral in Eq. \eqref{marginalization}, the one that includes the effective quantities, is given by 
\begin{multline}\label{eq:prior_effective}
    p(M_1^{z,\mathrm{eff}},D_L^\mathrm{eff}|z_\mathrm{c}, r, \theta, M_1, \mathcal{H}_A) =\\ \delta\qty(D_L^\mathrm{eff}-\qty(1+z_\mathrm{rel}(r,\theta))^2\qty(1+z_\mathrm{grav}(r))D_L(z_\mathrm{c},\Omega))\times \\ \times \delta\qty(M_1^{z,\mathrm{eff}} - \qty(1+z_\mathrm{rel}(r,\theta))\qty(1+z_\mathrm{grav}(r))\qty(1+z_c)M_1)\,.
\end{multline}
The expressions for $z_\mathrm{rel}$ and $z_\mathrm{grav}$ are the ones in Eqs.~\eqref{eq:relrs} and~\eqref{eq:grars} respectively, while $D_L(z_\mathrm{c},\Omega)$ denotes the luminosity distance associated with redshift $z_\mathrm{c}$ under the assumption of a set of cosmological parameters $\Omega$. Here we make use of the values reported in \citet{aghanim:2021}.

The common origin hypothesis for both the GW and EM transients fixes the cosmological redshift:
\begin{equation}
    p(z_\mathrm{c}|\mathcal{H}_A) = \delta(z_\mathrm{c} - z_\mathrm{AGN})\,.
\end{equation}
The prior for the distance from the SMBH $p(r|\mathcal{H}_A)$ is based on 
{the prediction of migration traps \cite{bellovary:2016} in accretion disk models by Sirko and Goodman \citet{sirko:2003}. We adopt the model with a constant, high accretion rate (Eddington ratio of $0.5$)}, 
as well as a linear dependence on radial distance to account for the overall axial symmetry of the system. Migration trap locations at 24.5 and 331 Schwarzschild radii\footnote{{The precise location of migration traps will vary depending on the disk model \citep{derdzinski:2023}, but we find this does not strongly affect our results.}} are approximated as Laplacian resonances centered at the migration trap locations with a full width at half maximum of 4.8 $R_s$.
Concerning the viewing angle $\theta$, we assume $p(\theta|\mathcal{H}_A)$ to be uniform in $\cos(\theta)$.

\begin{figure}
    \centering
    \includegraphics[width=\columnwidth]{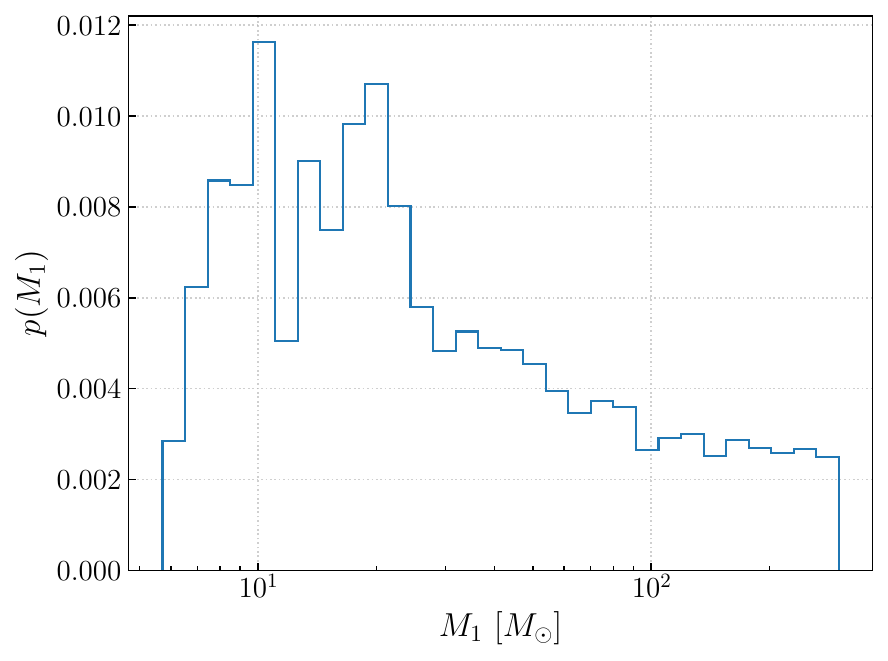}
    \caption{Primary mass probability density in the range $[0, 300]\ \msun$ of BBHs formed in the migration trap of AGN disks \citep{vaccaro:2023}.}
    \label{fig:AGN_mas_distr}
\end{figure}

The primary mass distribution prior $p(M_1|\mathcal{H}_A)$ follows the mass distribution presented in \citet{vaccaro:2023} for BBHs formed in 
AGN disks (see Fig.~\ref{fig:AGN_mas_distr}). It displays two peaks at roughly $10\ \msun$ and $20\ \msun$, populated by stellar-origin BHs, and a high-mass tail extending up to $10^4\ \msun$ populated by dynamically-formed BHs. The posterior distribution for the primary mass of GW190521 in GWTC-2.1 has support only below roughly $160\ \msun$, hence we limit our range of interest to the interval $\left[0,300\right]\ \msun$.

\subsubsection{Coincidence model}
In the coincidence model $\mathcal{H}_C$, the EM counterpart candidate and GW190521 are not associated and appear within the same sky localization by chance. Since the coincidence model for the GW event does not make use of any information from the EM counterpart candidate, the sky position is not conditioned on the sky location of the AGN. We then assume that $p(\alpha,\delta|\mathcal{H}_C)$ is the same uniform prior used for PE.

In this model, the only redshift present is the cosmological one, therefore the prior for the intrinsic primary mass and luminosity distance reads
\begin{multline}
    p(M_1^z,D_L|\mathcal{H}_C) = \int p(M_1^z|M_1,z_\mathrm{c},\mathcal{H}_C)p(D_L|z_\mathrm{c},\mathcal{H}_C)\times \\ \times p(M_1|\mathcal{H}_C) p(z_\mathrm{c}|\mathcal{H}_C)\dd z_\mathrm{c}\dd M_1\,.
\end{multline}
Under the hypothesis $\mathcal{H}_C$, the two probability densities now read
\begin{equation}
    p(M_1^z|M_1,z_\mathrm{c},\mathcal{H}_C) = \delta\qty(M_1^z-(1+z_\mathrm{c})M_1)\,,
\end{equation}
and
\begin{equation}
    p(D_L|z_\mathrm{c}, \mathcal{H}_C) = \delta\qty(D_L-D_L(z_\mathrm{c},\Omega))\,.
\end{equation}
Consequently, the only parameters of this model are the source-frame primary mass $M_1$ and the cosmological redshift $z_\mathrm{c}$ of the source.
The prior on the primary mass $p(M_1|\mathcal{H}_C)$ follows the \textsc{PowerLaw+Peak} model using the median parameters quoted in \citet{astrodist:gwtc3}, which includes the mass gap at the observed masses, whereas the prior distribution for $z_\mathrm{c}$ is proportional to the comoving volume element:
\begin{equation}
    p(z_\mathrm{c}|\mathcal{H}_C) \propto \frac{\dd V_\mathrm{c}}{\dd z}\,.
\end{equation}

\subsection{GW posterior probability density}\label{sec:codes}
The LVK collaboration released, for every GW event contained in their catalogs, a set of samples\footnote{Publicly available via GWOSC: \url{https://gwosc.org}} drawn from their PE posterior distribution.
In this work, we make use of the posterior samples released within GWTC-2.1 drawn from the 4-dimensional posterior distribution $p(M_1^z,D_L,\alpha,\delta|d)$ making use of the IMRPhenomXPHM waveform. This waveform model is the only one included in GWTC-2.1 for GW190521 \citep{gwtc2.1}: therefore, while comparing our findings with the existing literature, we will quote the results obtained with the waveform belonging to the same family, IMRPhenomPv3HM.

These samples are used to infer an analytical approximant using a Dirichlet process Gaussian mixture model (DPGMM), which is a flexible approximant based on an infinite mixture of Gaussian distribution capable of reconstructing arbitrary multivariate probability densities \citep{nguyen:2020}. Making use of this approximant, we are able to analytically condition and/or marginalize the posterior distribution $p(M_1^z,D_L,\alpha,\delta|d)$ on some of its variables ($\alpha$ and $\delta$, in this case).

We obtained the approximant for $p(M_1^z,D_L,\alpha,\delta|d)$ making use of \textsc{figaro}\footnote{Publicly available at \url{https://github.com/sterinaldi/FIGARO}.} \citep{rinaldi:2022}, a \textsc{Python} code designed to infer probability densities with a DPGMM. The integral in Eq.~\eqref{eq:likelihood_generic} is evaluated with \textsc{RayNest}\footnote{Publicly available at \url{https://github.com/wdpozzo/raynest}.}, a \textsc{Cython}-based implementation of the nested sampling algorithm, which is a sampling scheme primarily designed for evidence computation \citep{skilling:2006}.

\section{Results}
\label{Results: association model}
\begin{figure} 
    \centering
    \includegraphics[width=\columnwidth]{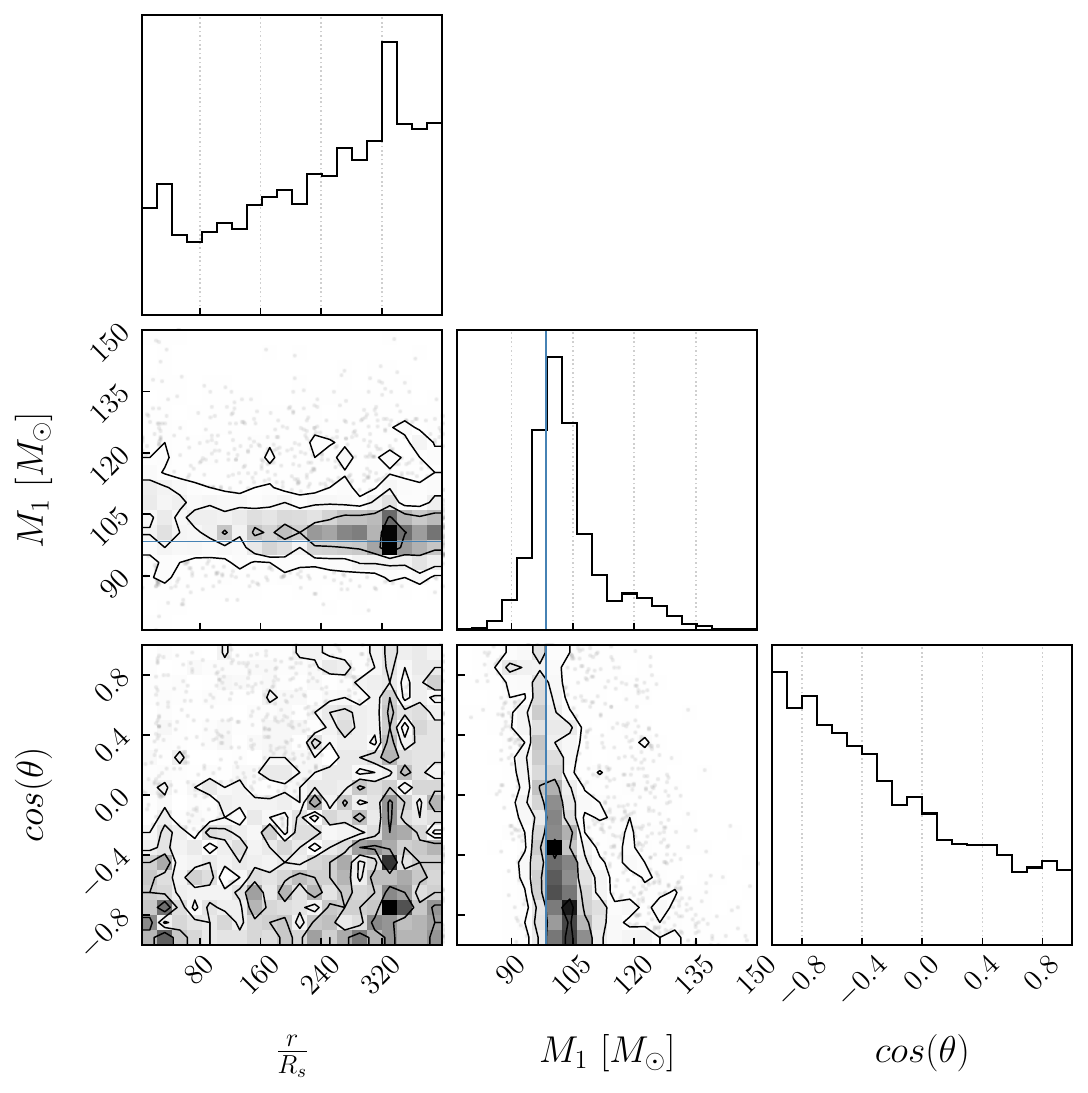}
  \caption{Posterior samples for the association model. The vertical blue line marks GW190521's median primary mass as reported in GWTC-2.1, $98.4^{+33.6}_{-21.7} \ \msun$}
  \label{joint_posterior}
\end{figure}

Figure~\ref{joint_posterior} shows the posterior distribution for the parameters of the association model: the radial distance from the central SMBH is consistent with the prior described above, other than slight deviations in the innermost region ($\lesssim 10 \ R_s$). This implies that the effect of the gravitational redshift is negligible.
The BBH motion towards the observer is favored, with a cosine of the orbital angle equal to -1 defined as motion directly towards the Earth. This motion corresponds to a blueshift of the GW, leading to a detector frame mass that is smaller than the corresponding mass for a BBH at rest in the comoving frame. 

\begin{figure}
    \centering
    \includegraphics[width=\columnwidth]{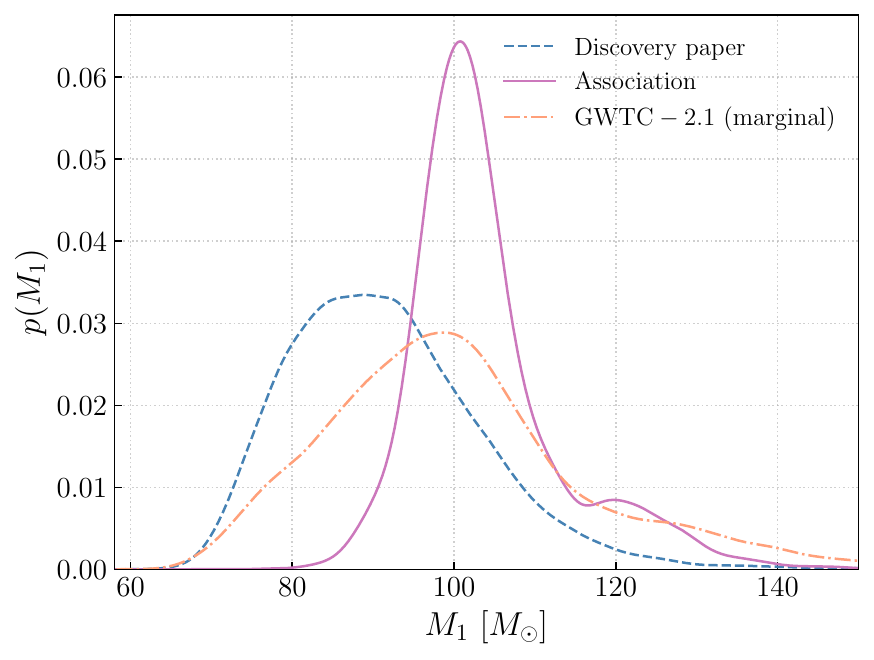}
    \caption{Posterior distribution for the source frame primary mass for the association model (solid pink) compared to the discovery paper (dashed blue) and the more recent GWTC-2.1 (dot-dashed orange) posterior distributions.}
  \label{mass_posterior}
\end{figure}

Figure~\ref{mass_posterior} compares different primary mass posterior probability densities: the one obtained with the association model is slightly shifted towards higher masses and is better constrained compared to the GWTC-2.1 distribution. Both distributions favor a more massive object with respect to the discovery paper.
Making use of the nested sampling algorithm, we are able to compute the evidence for each of the two models, thus making it possible to do model selection. The Bayes' factor favors the common origin hypothesis: $\log \mathcal{B}^A_C = 8.6\pm 0.1$.
To be sure that this result is not driven by the different astrophysical priors $p(M_1|\mathcal{H}_i)$ and in particular by the absence of the mass gap in the AGN formation channel, we repeated the analysis with a variation of the \textsc{PowerLaw+Peak} model without the mass gap as a primary mass prior for the coincidence model, finding an analogous value for {the log Bayes' factor: $\log \mathcal{B}^A_C = 8.5\pm 0.1$. The Bayes' factor is driven predominantly by two factors, with similar contribution: the updated luminosity distance in GWTC-2.1 compared to GWTC-2 and the different mass priors adopted for the two models. We find that the additional redshift contribution to the Bayes' factor is minimal: this can be understood
from the fact that the posterior distribution for $r/R_\mathrm{s}$ in Fig.~\ref{joint_posterior} closely follows the prior probability density, thus suggesting that the available data are not informative in this regard.}

Odds ratios require astrophysical prior odds for the random association between an AGN flare and GW190521. When computing odds ratios, we adopt $\mathcal{P}^A_C = 1/13$ from \citet{ashton:2021} as the astrophysical prior odds for a common source, given as the inverse of potentially associated flares within certain spatial and temporal bounds. Within the ZTF data, the number of flares similar to ZTF19abanrhr within the time span was 13.
Even with the inclusion of this prior, which favors random coincidences, the association model is largely favored -- $\log \mathcal{O}^A_C = 6.0\pm 0.1$ -- confidently relating the EM signal with the GW event.

\section{The Hubble Constant}\label{Hubble Constant}
GW events with EM counterparts can be used to determine the Hubble constant $H_0$ and ease the Hubble tension \cite{di-valentino_mena_2021,dainotti_de-simone_2021,perivolaropoulos_skara_2021,jedamzik_pogosian_2021,vagnozzi_2021,krishnan_mohayaee_2021,freedman_2021,di-valentino_anchordoqui_2021,rameez_sarkar_2021,dainotti_de-simnoe_2022,vagnozzi_pacucci_2022,mortsell_goobar_2022}. 
To date, only one GW event, the binary neutron star (BNS) merger GW170817 \citep{gw170817:2017}, has been confidently associated with its EM counterpart \citep{gw170817:2017:cosmology}, thus allowing for a \emph{standard siren measurement} of the Hubble constant \citep{gw170817:2017:cosmology}.

Both \citet{mukherjee:2021} and \citet{chen:2022}, following the potential association of \citet{ashton:2021}, analyzed GW190521 as a standard siren. The former finds a value of $H_0 = 43.1^{+24.6}_{-11.4}\mathrm{\ km \ s^{-1} \ Mpc^{-1}}$ using the discovery paper IMRPhenomPv3HM posterior samples\footnote{In \citet{mukherjee:2021}, the authors consider also two alternative waveform models not included in the GWTC-2.1 data release, NRSur7dq4 and SEOBNRv4PHM, finding $H_0 =62.2^{+29.5}_{-19.7}\mathrm{\ km \ s^{-1} \ Mpc^{-1}}$ and $H_0=50.4^{+28.1}_{-19.5}\mathrm{\ km \ s^{-1} \ Mpc^{-1}}$ respectively.} and an uninformative prior on $H_0$. 
\citet{chen:2022} finds $H_0 = 48^{+23}_{-10}\mathrm{\ km \ s^{-1} \ Mpc^{-1}}$ making use of the discovery paper NRSur7dq4 posterior samples. {For both values, the associated uncertainty corresponds to the 68\% credible interval. Both of these works favors a lower value of $H_0$ compared to the SH0ES \citep{riess:2022} ($73.04 \pm 1.04\mathrm{\ km \ s^{-1} \ Mpc^{-1}}$) and Planck \citep{aghanim:2021} ($67.4 \pm 0.5 \mathrm{\ km \ s^{-1} \ Mpc^{-1}}$) values.}
\citet{gayathri:2020:cosmo}, using the same data set, considers the possibility of a highly eccentric binary, measuring $H_0 = 88.6^{+17.1}_{-34.3}\mathrm{\ km \ s^{-1} \ Mpc^{-1}}$.

We extend the statistical framework presented in Section~\ref{Bayesian statistical framework} to include the Hubble constant among the parameters of the association model and infer a probability density for $H_0$. In Eq.~\eqref{eq:prior_effective}, we make use of the cosmological parameters $\Omega$, including $H_0$ to convert the cosmological redshift into a luminosity distance. Instead of fixing the whole set of $\Omega$ parameters, we treat $H_0$ as a free parameter of our association model. The remaining values (density and dark energy equation of state parameters) are, again, taken from \citet{aghanim:2021}.

\begin{figure} 
    \centering
    \includegraphics[width=\columnwidth]{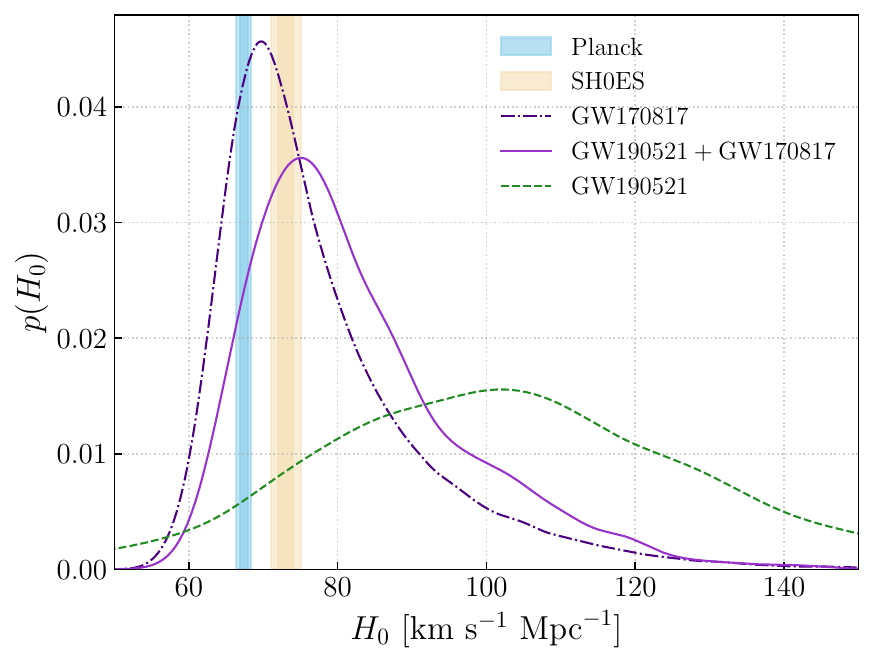}
    \caption{Hubble constant posterior distributions for GW170817, GW190521, and joint GW190521 + GW170817 compared to the reported values by Planck and SH0ES (1- and 2-$\sigma$ credible intervals).}
  \label{H_0}
\end{figure}

We explore two different prior choices for $H_0$ under the association model, with and without prior knowledge of the $H_0$ posterior distribution obtained with GW170817:
\begin{itemize}
 \item \textbf{Uninformative prior:} we assume no previous knowledge about $H_0$, particularly about its magnitude: we convey this via a flat-in-log prior: $p(H_0|\mathcal{H}_A) \propto 1/H_0$; 
  \item \textbf{GW170817 prior:} we use the GW170817 $H_0$ posterior (\citet{gw170817:2017:cosmology}, dot-dashed line in Fig.~\ref{H_0}) as a prior distribution. This is equivalent to jointly analyzing GW170817 and GW190521.
\end{itemize}

Figure~\ref{H_0} shows the posterior distributions for $H_0$ under the two different prior choices. 
With the agnostic, flat-in-log prior, we find $H_0 = 102^{+27}_{-25}\mathrm{\ km \ s^{-1} \ Mpc^{-1}}$, whereas, with a prior distribution that accounts for GW170817, $H_0$ is found to be $79.2^{+17.6}_{-9.6} \mathrm{\ km \ s^{-1} \ Mpc^{-1}}$. Both distributions are consistent with Planck \citep{aghanim:2021} and SH0ES \citep{riess:2022}, with a preference towards larger $H_0$ values. 

When compared to the coincidence model, the common origin hypothesis with the agnostic $H_0$ prior is preferred with a log Bayes' factor $\log \mathcal{B}^A_C = 8.5 \pm 0.1$, corresponding to $\log \mathcal{O}^A_C = 5.9 \pm 0.1$ when astrophysical prior odds are included. The GW170817-informed prior favors the association model even more, resulting in $\log \mathcal{B}^A_C = 9.1\pm 0.1$ and $\log \mathcal{O}^A_C = 6.6 \pm 0.1$.
The fact that our $H_0$ inference with an agnostic prior is consistent with the existing literature adds a piece of circumstantial evidence in favor of the association claimed in the previous Section.

\section{Conclusions}\label{conclusion}

In this work, we investigate the possibility of the AGN flare ZTF19abanrhr to be the EM counterpart of the BBH merger GW190521, thus identifying AGN 124942.3+344929 to be the host galaxy of this GW event. Our model accounts for the environmental effects due to the proximity of the BBH to the SMBH in the center of the AGN disk including two additional redshift terms.

Making use of the GWTC-2.1 \citep{gwtc2.1} posterior samples, the model in which the two transients share a common origin is preferred with respect to the random coincidence hypothesis, even assuming astrophysical prior odds of 1/13 \citep{ashton:2021} based on the \emph{a priori} possibility of a random coincidence. The Bayes' factors and odds ratios for all the models presented in Secs.~\ref{Results: association model} and~\ref{Hubble Constant} compared to the coincidence model are shown in Table~\ref{tab:Bayes}. 

\begin{table}
  \begin{center}
    \caption{Bayes' factors and odds ratios for the different models considered in this work compared to the coincidence model.}
    \begin{tabular}{lcc}
      \midrule
      \midrule
      \textbf{Model} & $\log \mathcal{B}^A_C$ & $\log \mathcal{O}^A_C$\\
      \midrule
      Association, fixed $H_0$ & 8.6 $\pm$ 0.1 & 6.0 $\pm$ 0.1\\
      Free $H_0$, uninformative prior & 8.5 $\pm$ 0.1 & 5.9 $\pm$ 0.1 \\
      Free $H_0$, GW170817 prior & 9.1 $\pm$ 0.1 & 6.6 $\pm$ 0.1 \\
      \midrule
      \midrule
    \end{tabular}
    \label{tab:Bayes}
  \end{center}
\end{table}

The additional redshift of the BBH caused by its orbit around the AGN's SMBH results in a slightly different, albeit consistent, source frame primary mass with respect to the coincidence model. We find that the GW is likely to be blueshifted, increasing the source frame mass to $101.9^{+9.5}_{-5.4} \ \msun$ compared to the reported GWTC-2.1 value of $98.4^{+33.6}_{-21.7} \ \msun$. 

Making use of the bright siren method, we report a measurement of the Hubble constant $H_0 = 102^{+27}_{-25}\ \mathrm{km \ s^{-1} \ Mpc^{-1}}$. Additionally, the independent multi-messenger detection of the BNS merger GW170817 can be used as informed prior for $H_0$. In this case, $H_0$ is found to be $79.2^{+17.6}_{-9.6}\ \mathrm{km \ s^{-1} \ Mpc^{-1}}$. The two distributions, although consistent with both Planck \citep{aghanim:2021} and SH0ES \citep{riess:2022}, hint towards a larger value of $H_0$, contrary to the findings of \citet{mukherjee:2021} and \citet{chen:2022}.

While previously puzzling for the expectation of a mass gap above $80\ \msun$, our findings that GW190521 is likely to have originated in an AGN helps to reconcile the observation of this massive binary system with the existing astrophysical models.

\begin{acknowledgments}
The authors are grateful to Michela Mapelli, Juan Calderón Bustillo and Samson Leong for the useful discussion.

SM acknowledges support by the University of Florida's International Research Experience for Undergraduates program, funded by the NSF (Grant agreement NSF PHY-1950830).
SR acknowledges support by the EU Horizon 2020 Research and Innovation Program under the Marie Sklodowska-Curie Grant Agreement No. 734303.

This research has made use of data or software obtained from the Gravitational Wave Open Science Center (gwosc.org), a service of the LIGO Scientific Collaboration, the Virgo Collaboration, and KAGRA. This material is based upon work supported by NSF's LIGO Laboratory which is a major facility fully funded by the National Science Foundation, as well as the Science and Technology Facilities Council (STFC) of the United Kingdom, the Max-Planck-Society (MPS), and the State of Niedersachsen/Germany for support of the construction of Advanced LIGO and construction and operation of the GEO600 detector. Additional support for Advanced LIGO was provided by the Australian Research Council. Virgo is funded, through the European Gravitational Observatory (EGO), by the French Centre National de Recherche Scientifique (CNRS), the Italian Istituto Nazionale di Fisica Nucleare (INFN) and the Dutch Nikhef, with contributions by institutions from Belgium, Germany, Greece, Hungary, Ireland, Japan, Monaco, Poland, Portugal, Spain. KAGRA is supported by Ministry of Education, Culture, Sports, Science and Technology (MEXT), Japan Society for the Promotion of Science (JSPS) in Japan; National Research Foundation (NRF) and Ministry of Science and ICT (MSIT) in Korea; Academia Sinica (AS) and National Science and Technology Council (NSTC) in Taiwan.

Funding for the Sloan Digital Sky Survey (SDSS) has been provided by the Alfred P. Sloan Foundation, the Participating Institutions, the National Aeronautics and Space Administration, the National Science Foundation, the U.S. Department of Energy, the Japanese Monbukagakusho, and the Max Planck Society. The SDSS Web site is http://www.sdss.org/.

The SDSS is managed by the Astrophysical Research Consortium (ARC) for the Participating Institutions. The Participating Institutions are The University of Chicago, Fermilab, the Institute for Advanced Study, the Japan Participation Group, The Johns Hopkins University, Los Alamos National Laboratory, the Max-Planck-Institute for Astronomy (MPIA), the Max-Planck-Institute for Astrophysics (MPA), New Mexico State University, University of Pittsburgh, Princeton University, the United States Naval Observatory, and the University of Washington.
\end{acknowledgments}

\bibliography{bibliography}
\end{document}